\documentclass[twocolumn,prb]{revtex4-2}
\usepackage{epsfig,amssymb,amsmath}
\usepackage{graphicx}
\usepackage{color}
\usepackage[english]{babel}

\begin{document}

\title{Locking of magnetization and Josephson oscillations at ferromagnetic resonance in $\varphi_0$ junction under external radiation}
\author{S. A. Abdelmoneim~$^{1,2}$}
\author{Yu. M. Shukrinov~$^{1,3,4}$}
\author{K. V. Kulikov~$^{1,3}$}
\author{H. ElSamman~$^{2}$}
\author{M. Nashaat~$^{1,5}$}

\address{
$^{1}$ BLTP, JINR, Dubna, Moscow Region, 141980, Russia\\
$^{2}$ Physics department, Menofiya University,  Faculty of Science, 32511, Shebin Elkom,Egypt\\
$^{3}$ Department of Nanotechnology and New Materials, Dubna State University, Dubna,  141980, Russia\\
$^{4}$ Moscow Institute of Physics and Technology, Dolgoprudny, 141700, Moscow Region, Russia\\
$^{5}$ Department of Physics, Faculty of Science, Cairo University, 12613, Giza, Egypt
}

\date{\today}

\begin{abstract}
We demonstrate the locking by external electromagnetic radiation of magnetic precession in the $ \varphi_0 $ Josephson junction through Josephson oscillations in the region of ferromagnetic resonance. This leads to a step in the dependence of the magnetization on the bias current, the position of which is determined by the radiation frequency and the shape of the resonance curve. In junctions with a strong spin-orbit coupling, states with negative differential resistance appear on the IV-characteristic, resulting in an additional locking step. A detailed study of the time dependence of voltage and magnetization and their Fourier transforms shows that the corresponding oscillations have the same frequency as the oscillations at the first step, but they have a different amplitude and different dependence on the radiation frequency. This makes it possible to control not only the frequency, but also the amplitude of the magnetic precession in the locking region. It opens up unique perspectives for the control and manipulation of magnetic moment in hybrid systems such as anomalous Josephson junctions.

\end{abstract}

\maketitle

\paragraph*{ Introduction.}
The $\varphi_0$-junction related to a special class of anomalous Josephson structures with coupled superconducting and magnetic characteristics allows  to manipulate  the magnetic  properties by Josephson current ~\cite{linder15,efetov11,buzdin05,bergeret05,golubov04,ghosh17}. It demonstrates a number of interesting features important for superconducting spintronics and modern informational technologies \cite{yokoyama14,minutillo18,krive05,reynoso08,alidoust17,alidoust18,braude07,zyuzin16,zyuzin00,alidoust18-2,goldobin11,goldobin15,menditto18,alidoust13,shapiro18,spanslatt18}. The current-phase relation of the $\varphi_0$-junction is given by $I = I_c \sin (\varphi-\varphi_0)$, where the phase shift $\varphi_0$ is proportional to the magnetic moment perpendicular to the gradient of the asymmetric spin-orbit potential~\cite{buzdin08,konschelle09}. Supercondoctor spintronics, based on  anomalous Josephson junctions,  has evolved into a major field of research that broadly encompasses different classes of materials, magnetic systems, and devices~\cite{linder15,eschrig18,Ferr,Igor,J. R. Hauptmann,liu10-1,liu10-2,konschelle15}.

One of the challenges for superconductor electronics and spintronics which stands out by ultra-low  energy dissipation is the creation of cryogenic memory \cite{nguyen2019,bergeret19,herr11,baek14,mukhanov11,birge15,apl17}. In Ref.\ \cite{mazanik20} an analytical solution for the magnetization dynamics induced by an arbitrary current pulse  and  the criteria for magnetization reversal in the $\varphi_0$ junction have been proposed. The obtained results explain  the periodicity in the appearance of the magnetization reversal intervals observed in Ref.\ \cite{jetpl-atanas19} and allow one to predict magnetization reversal at the chosen system parameters.

Recently, an anomalous phase shift was experimentally observed in different systems, particularly, in the $\varphi_0$  junction based on a nanowire quantum dot \cite{szombati16} and directly through CPR measurement in a hybrid SNS JJ fabricated using ${\rm Bi_2Se_3}$ (which is a topological insulator with strong spin-orbit coupling) in the presence of an in-plane magnetic field~\cite{aprili19}. The observation of a tunable anomalous Josephson effect in InAs/Al Josephson junctions measured via a superconducting quantum interference device (SQUID) reported in Ref.\ \cite{mayer19}. The authors were able to tune the spin-orbit coupling of the Josephson junction by more than one order of magnitude. This gives  the ability to tune  $\varphi_0$, and opens several new opportunities for superconducting spintronics \cite{linder15}, and new possibilities for realizing and characterizing topological superconductivity \cite{alicea12,fornieri19,ren19}. In Ref.\ \cite{chudn2016,chudn2010}, the authors argued that the $\varphi_0$  Josephson junction is ideally suited for studying of quantum tunneling of the magnetic moment. They proposed that magnetic tunneling would show up in the ac voltage across the junction and it could be controlled by the bias current applied to the junction.

Though the static properties of the SFS structures are well studied both theoretically and experimentally, much less is known about the magnetic dynamics of these systems~\cite{waintal02,braude08,linder83,holmqvist14,holmqvist18,bergeret18,eschrig18,mironov21,villas21,lu20,ojajarvi20,hamar16,abdollahipour15} Different type of very simple  and harmonic  precessions of the magnetic moment were demonstrated in Ref.\ \cite{shukrinov-prb19}. Such  magnetic precessions may be monitored by the appearance of higher harmonics in the current-phase relation and by the presence of a dc component of the superconducting current \cite{konschelle09}. It is expected that external radiation would lead to a series of novel phenomena. The possibility of appearance  of half-integer Shapiro steps, in addition to the conventional integer steps,  and  the generation of an additional magnetic precession with frequency of external radiation was already discussed in Ref.\ \cite{konschelle09}.

Nonlinear superconducting structures being driven far from equilibrium exhibit a negative differential resistance (NDR) in the current-voltage characteristics \cite{Kleiner2008,Filatrella2014}. The NDR plays an essential role in many applications, in particular, for THz radiation emission \cite{Kadowaki2008,Lin2013}, implementing single devices with a two or more threshold voltage which provides electrically stable operating points during the circuit operation \cite{Lin2015,Choi2015,Campbell2015,Roy2015,Shim2016,Campbell2016, Kobashi2018, Kim2020}. Here we demonstrate an important role of the states with NDR in the locking of magnetization and Josephson oscillations at ferromagnetic resonance (FMR) in $\varphi_0$ junction under external radiation.

In this paper the IV characteristics and magnetization dynamics of the $\varphi_0$ Josephson junction under an external electromagnetic radiation are studied. We solve system of Landau-Lifshitz-Gilbert-Josephson equations  which takes into account the interaction of Josephson oscillations and magnetic moment of ferromagnetic layer. The bias current dependence of maximal magnetization component $m_{y}^{max}(I)$ (taken at each value of current) manifests two phenomena, such as FMR and locking of the magnetization precession  to the oscillations of the external field through the locking to the Josephson oscillations.  The locking is manifested as a step on $m_{y}^{max}(I)$ dependence and  its maximum shows the FMR. We clarify the role of spin-orbit coupling  in the appearance of the nonlinearity in the IV-curve and additional Shapiro step (SS) at  small radiation amplitudes. We show that in junctions with a strong spin-orbit coupling the states with NDR result in an additional step with corresponding oscillations having the same frequency as the oscillations at the first step, but a different amplitude and  different dependence on the radiation frequency. This open a way to control not only the frequency, but also the amplitude of the magnetic precession in the locking region.  Unique perspectives appear for the control and manipulation of magnetic moment in hybrid systems such as anomalous Josephson junctions.

\paragraph*{  Model and Method. }

\begin{figure}[h!]
\centering
\includegraphics[width=6.7cm]{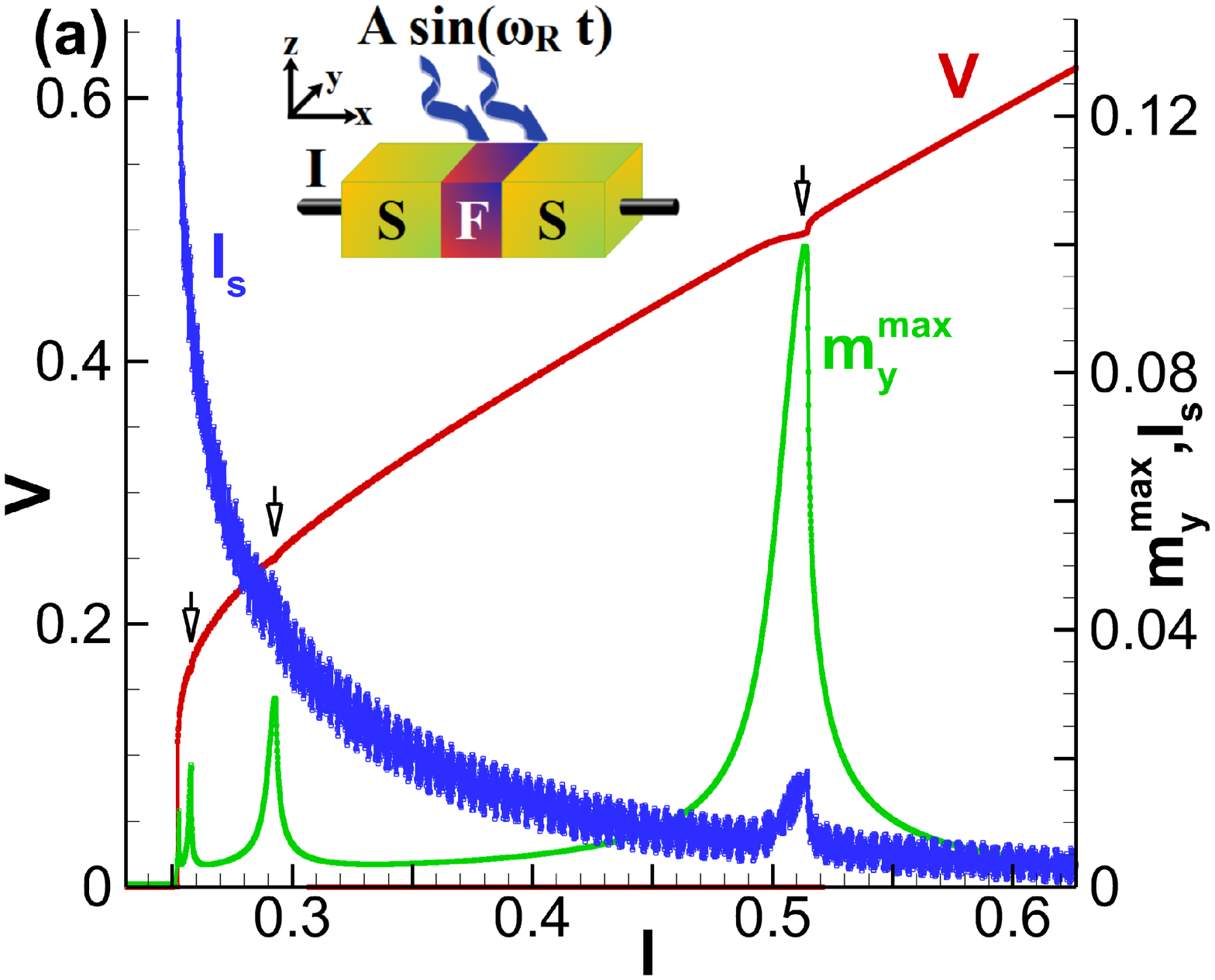}
\includegraphics[width=3.9cm]{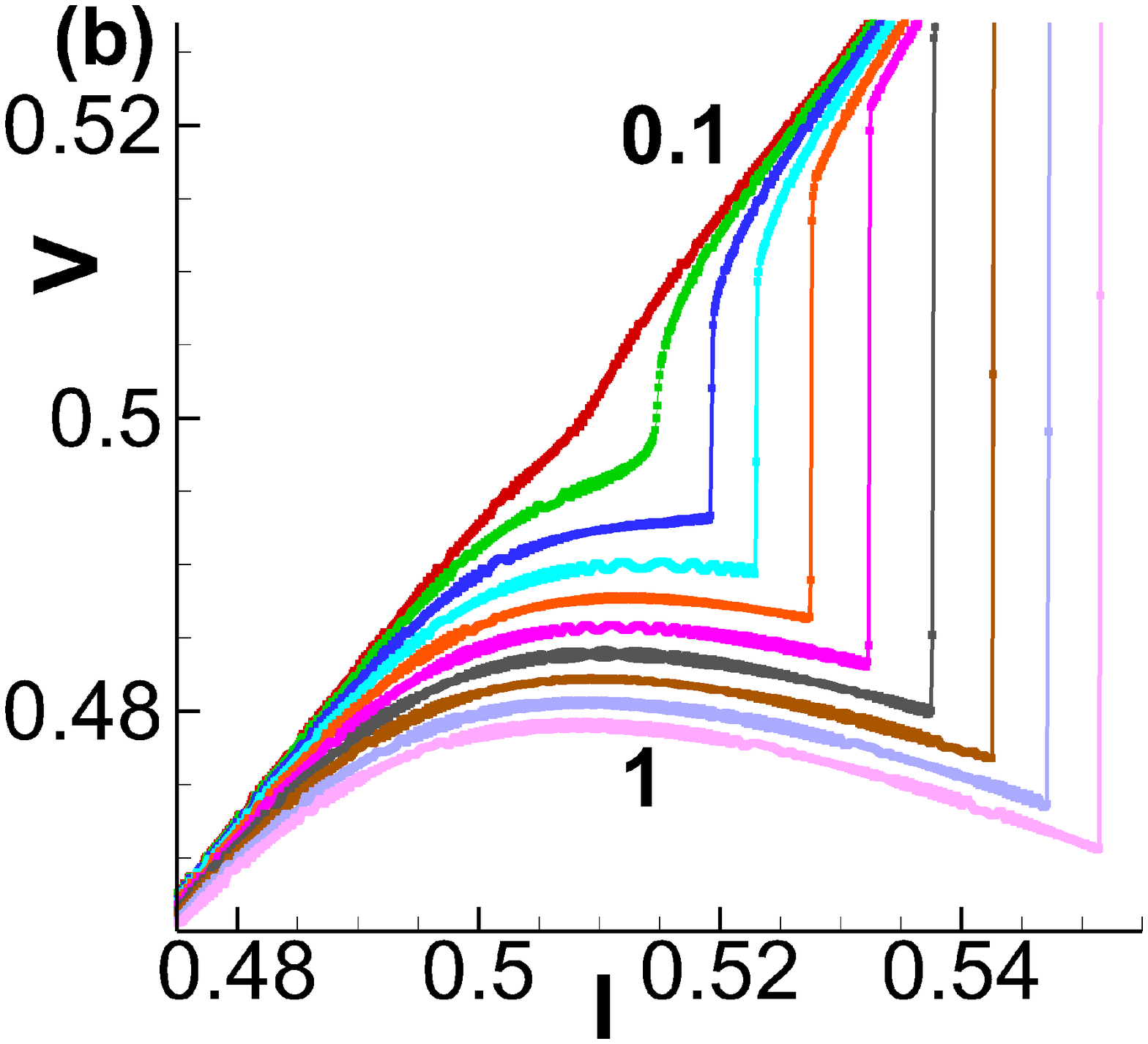}\includegraphics[width=4cm]{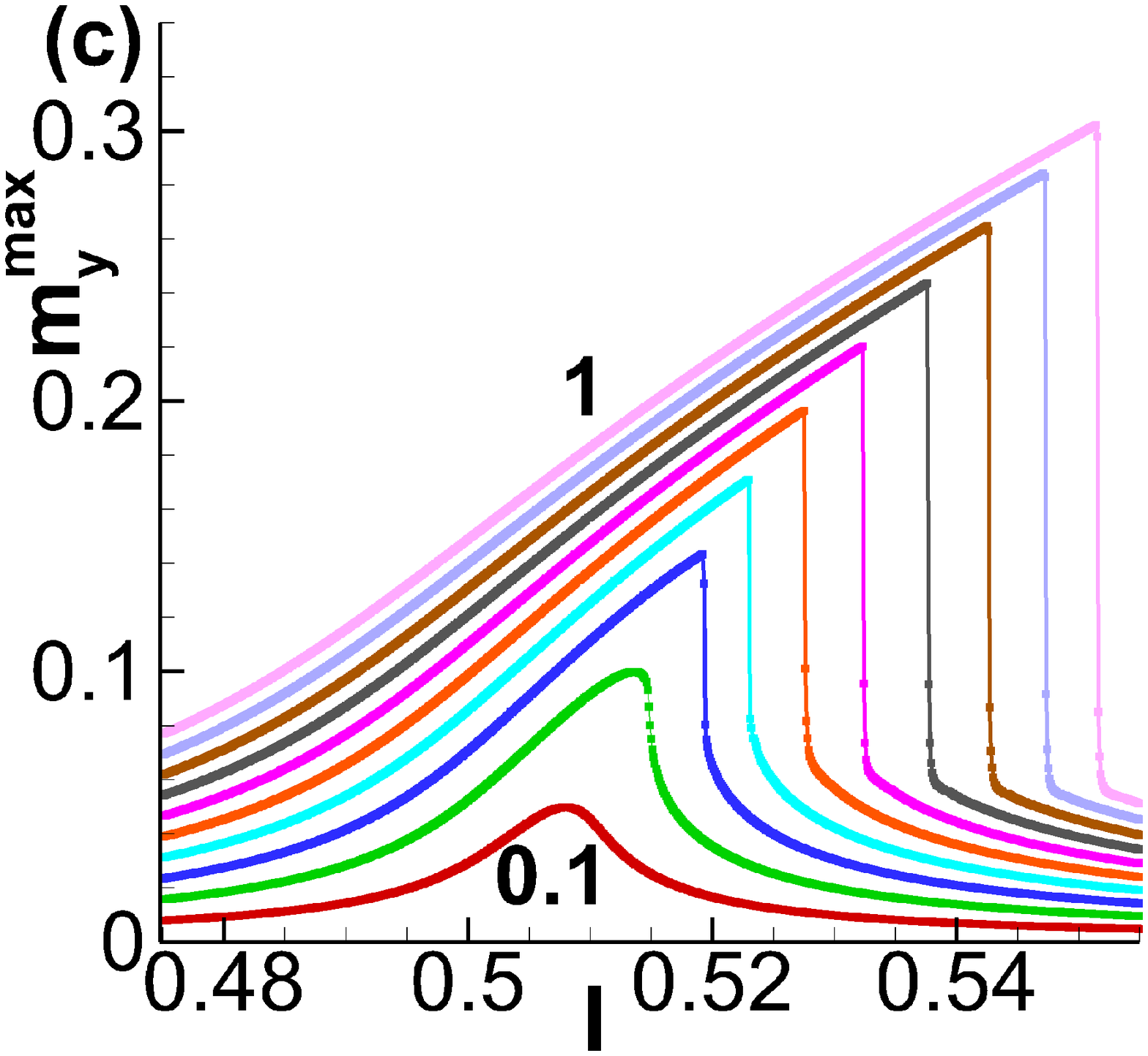}
\caption{(a) Manifestation of the FMR  in  $m_{y}^{max}(I)$, IV-characteristic and $ I_{s}(I)$ for the $\varphi_0$ junction which geometry is shown in the inset with S - superconductor, F - ferromagnet. Simulation parameters were  $G=0.01$, $r=0.2$ and $\alpha=0.01$; (b) Enlarged parts of IV-curves in the resonance region at different $r$. The numbers indicate the increasing of spin orbit coupling from $0.1$ to $1$ by an increment $0.1$; (c)  The same for $m_{y}^{max}(I)$.} \label{1}
\end{figure}

The geometry of the considered $\varphi_{0}$ junction is shown in Fig.\ref{1}(a). The ferromagnet easy-axis and the gradient of the spin-orbit potential (n) are directed along the z-axis.
In Josephson junctions with a thin ferromagnetic layer the superconducting phase difference and magnetization of the $F$ layer are two coupled dynamical variables. The system of equations describing the  dynamics of these variables is obtained from the Landau-Lifshitz-Gilbert (LLG) equation and Josephson relations for current and phase difference.

The total system of Landau-Lifshitz-Gilbert-Josephson equations \cite{konschelle09} (to be used in our numerical studies) in
normalized units  is given by:
\begin{equation}
\label{syseq}
\begin{array}{llll}
\displaystyle \dot{m}_{x}=\frac{\omega_F}{1+\alpha^{2}}\{-m_{y}m_{z}+Grm_{z}\sin(\varphi -rm_{y})\\
-\alpha[m_{x}m_{z}^{2}+Grm_{x}m_{y}\sin(\varphi -rm_{y})]\},
\vspace{0.2 cm}\\
\displaystyle \dot{m}_{y}=\frac{\omega_F}{1+\alpha^{2}}\{m_{x}m_{z}\\
-\alpha[m_{y}m_{z}^{2}-Gr(m_{z}^{2}+m_{x}^{2})\sin(\varphi -rm_{y})]\},
\vspace{0.2 cm}\\
\displaystyle \dot{m}_{z}=\frac{\omega_F}{1+\alpha^{2}}\{-Grm_{x}\sin(\varphi -rm_{y})\\
-\alpha[Grm_{y}m_{z}\sin(\varphi -rm_{y})-m_{z}(m_{x}^{2}+m_{y}^{2})]\},
\vspace{0.2 cm}\\
 \frac{d V}{d t}=\frac{1}{\beta_{c}}[I+ A \sin(\omega_R t)-V-\sin(\varphi-rm_{y})+r dm_{y})],\\
 \quad  \frac{d \varphi}{d t}=V,
 \end{array}
\end{equation}
\noindent where $\beta_{c}=2e I_{c}CR^{2}/\hbar$ is the McCumber
parameter (in our calculations we use $\beta_{c}=25$), $\varphi$ is the phase difference between the superconductors across the junction, $\displaystyle G= E_{J}/(K \mathcal{V})$, $K$ is an anisotropic
constant, $\mathcal{V}$ is the volume of ferromagnetic $F$ layer,
$r=l\upsilon_{so}/\upsilon_{F}$ is parameter of spin-orbit coupling, $\upsilon_{so}/\upsilon_{F}$ characterizes a relative
strength of spin-orbit interaction, $\upsilon_{F}$ is Fermi velocity, $l=4h L/\hbar \upsilon_{F}$, $L$ is the length of the $F$-layer, $h$ denotes the exchange field in the ferromagnetic layer, $\alpha$ is a phenomenological damping constant, $m_{i}= M_{i}/M_0$ for $i=x,y,z$, $M_{0}=\|{\bf M}\|$, $\omega_F=\Omega_F / \omega_c$ with the FMR frequency $\Omega_F=\gamma K/M_0$, $\gamma$ is the gyromagnetic ratio, characteristic frequency
$\omega_{c} =2eRI_{c}/\hbar$, $A$ is the amplitude of external radiation normalized to $I_c$ and $\omega_R$ is the frequency of external radiation normalized to $\omega_c$.
Here we normalize time in units of $\omega^{-1}_{c}$, external
current $I$ in units of $I_{c}$, and the voltage $V$ in units of
$V_{c}=I_{c}R$. This system of equations, solved numerically using
the fourth order Runge--Kutta method, yields $m_{i}(t)$, $V(t)$ and $\varphi(t)$ as a function of the
external bias current $I$. After averaging procedure~\cite{prb07,kleiner-book} we can find IV-characteristics at fixed system's parameters.

\paragraph*{ FMR and effect of the spin-orbit coupling}
Let us first discuss effect of spin-orbit coupling on IV-characteristics and magnetization in our system. FMR in $\varphi_0$ JJ is demonstrated in Fig.\ \ref 1(a), where we see an increase of magnetization amplitude $m_{y}^{max}$ (maximal $m_y$, calculated at each value of the bias current) in the resonance region near $\omega_F=0.5$. This resonance is also manifested in the IV-characteristics as the corresponding resonance branch shown by arrow  in this figure. We note  that due to the nonlinearety in our system, which reflects the nonlinear nature of the LLG equation, the resonance frequency decreases with an increasing in SOC or damping in the system, i.e. the resonance realized at $ \omega_J< \omega_F $ \cite{shukrinov-prb21}. So, the end of the resonance branch does not coincide with $\omega_F$. We see also the manifestation of two FMR subharmonics corresponded to  $V = \omega_F/2$ and $V = \omega_F/3$. The superconducting current which is demonstrated in this figure reflected  the FMR also \cite{shukrinov-prb19}.

An increase in the spin-orbit coupling at small $G$ and $\alpha$, when the nonlinearity in LLG is getting stronger, leads also to the manifestation of nonlinearity in the IV-characteristic. It has a pronounced effect on the shape of the IV curve in the resonance region, represented as the deviation of the IV curve from its linear behaviour and in the appearance of the the resonance branch as shown in Fig.\ \ref 1(b). A clear manifestation of a state with a negative differential resistance appears at $r>0.4$.

An increase in the spin-orbit interaction increases  also the peak in $m_{y}^{max}(I)$ dependence  and shifts it to the larger values of bias current as it is shown in Fig.\ \ref 1(c). It's interesting to note, that there is no any clear indications of the transformation to the state with negative differential resistance in this dependence. But, as we see below,  such manifestations appear  under the external radiation.

\paragraph*{ Effect of the external radiation on the S/F/S $\varphi_0$ JJ.}
Due to Gilbert damping, the FMR frequency is decreased in compare with a case at $\alpha=0$, so to observe the  effect of external radiation on IV-characteristics in FMR region, we apply the radiation with a smaller frequency than $\omega_F=0.5$ \cite{shukrinov-prb21}. The results of calculations at $\omega = 0.485$ and amplitude $A=0.1$  are presented in Fig.\ \ref 2(a).  It shows the voltage, the magnetization amplitude $m_{y}^{max}$ and the superconducting current $ I_{s}$ ploted versus the bias current in its downward direction at a fixed value of spin-orbit coupling $r=0.2$. We use  the same parameters as in the case without radiation presented in Fig.\ \ref 1(a). We see that dependence  $m_{y}^{max}(I)$ demonstrates both phenomena, i.e., as the locking of magnetization precession  and the FMR.  The locking of Josephson oscillations to the radiation frequency is manifested as the Shapiro step in the IV-characteristic. The maximum of $m_{y}^{max}(I)$ dependence demonstrates the FMR. The step in this dependence  is actually a locking of magnetization precession  to the oscillations of the external field through the locked Josephson oscillations.
\begin{figure}[h!]
\centering
\includegraphics[width=4.3cm]{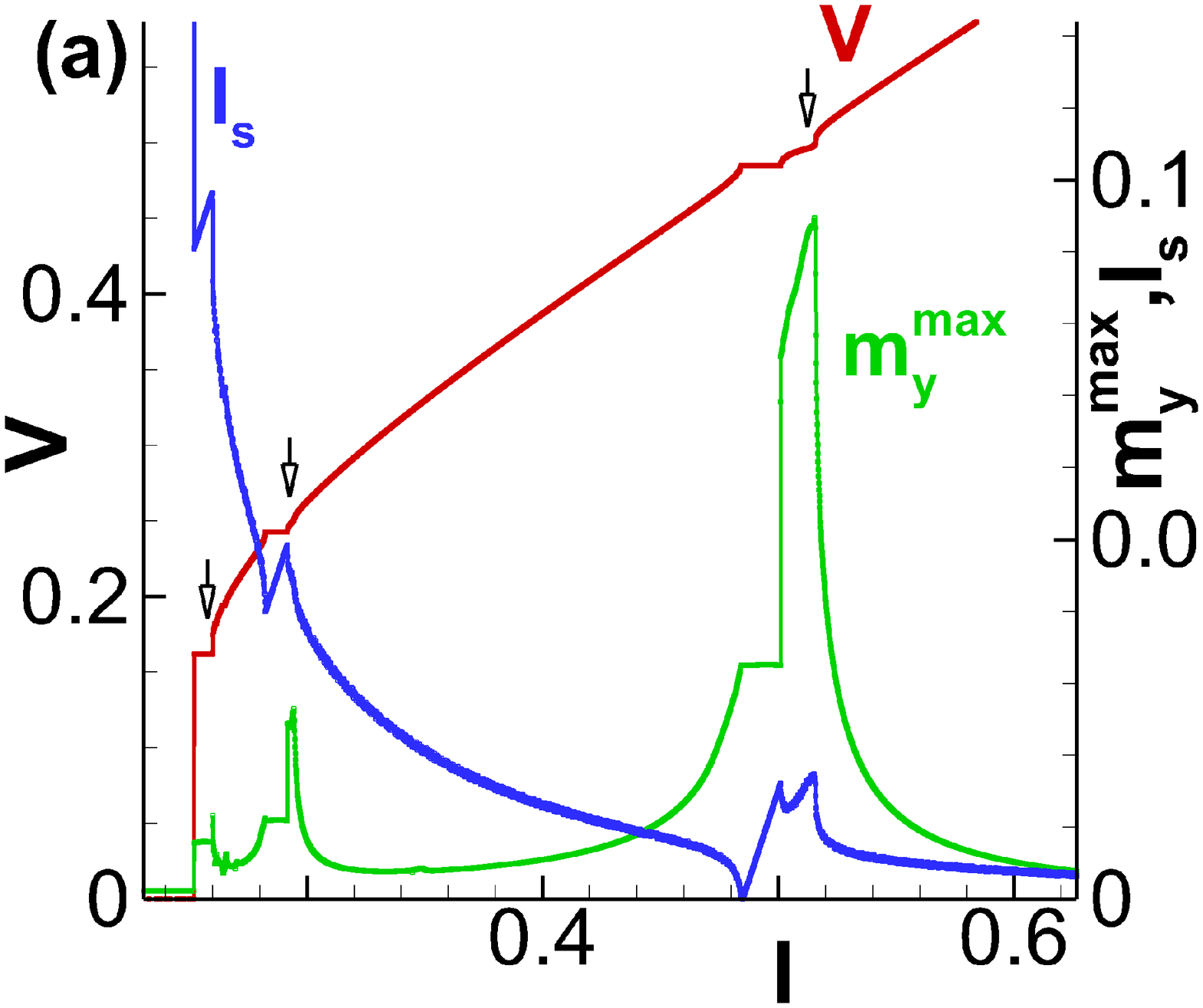}\includegraphics[width=4.3cm]{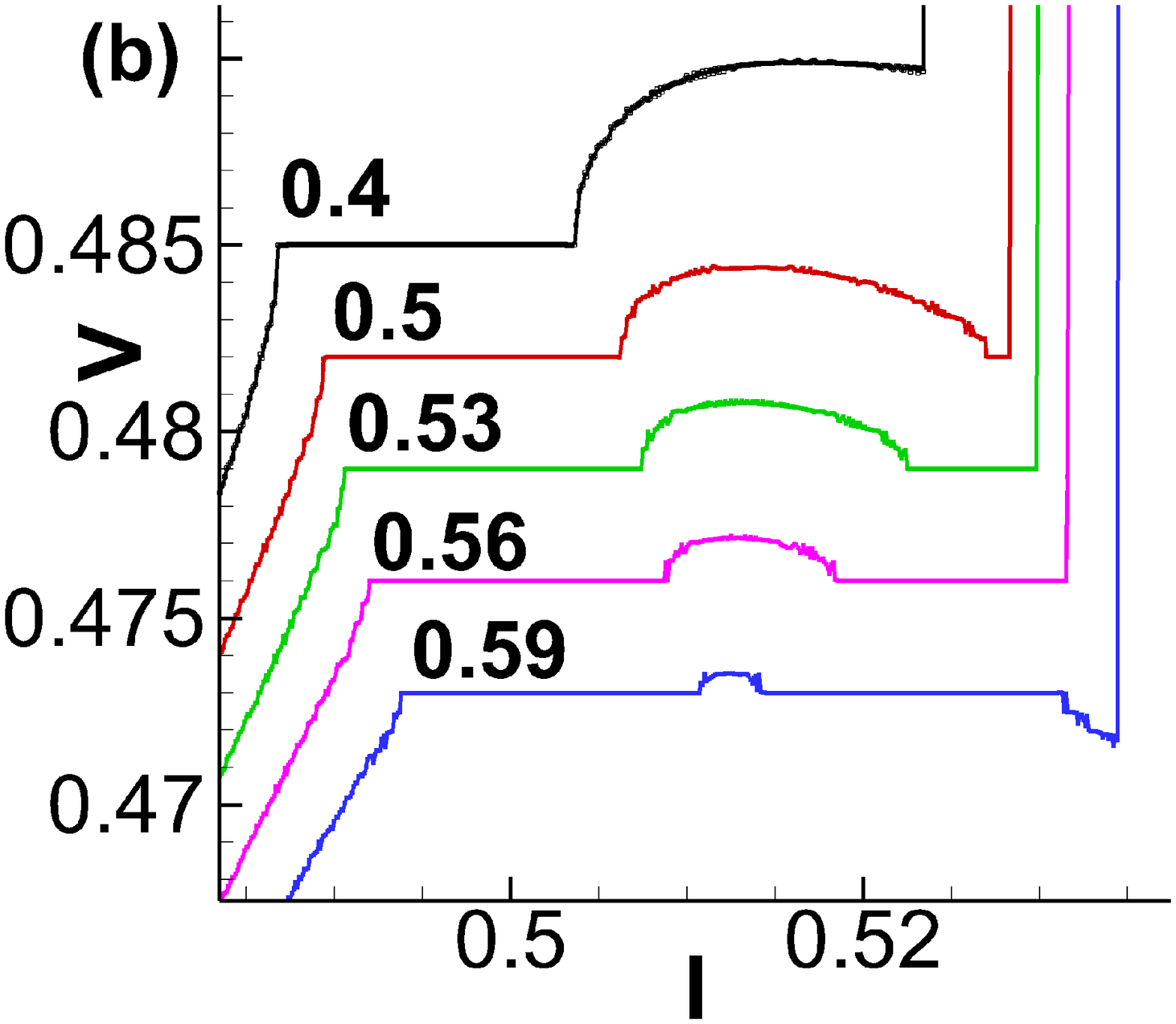}
\caption{(a) The same as in Fig.\ \ref 1(a) under radiation with $\omega_R=0.485$ and  $A=0.1$; (b)  An enlarged view of the IV-curves at different $r$. The curves are shifted down relatively the curve at $r=0.4$ by $\Delta V= 0.003$.  } \label{2}
\end{figure}
The average superconducting current obtained in during the same numerical simulations demonstrates a specific feature of Shapiro step indicated by the solid circle in Fig.\ \ref 2(a). So, $V(I)$, $m_{y}^{max}(I)$ and  $ I_{s}(I)$  show clearly the features related to the FMR and locking of Josephson and magnetic oscillations to the oscillations of the  external electromagnetic radiation. 

Effect of external radiation on the IV-characteristics at different values of spin-orbit coupling is demonstrated in Fig.\ \ref 2(b) for fixed radiation frequency and amplitude. We see the Shapiro step, a hump (dome) as a manifestation of the FMR and the second SS appeared in the part of the resonance branch with NDR. We emphasise that the NDR state leads to a unique situation when two SS in the IV-characteristic coexist at the same frequency.

With an increase in the SOC, the maximum of the resonance curves presented in Fig.\ \ref 1(b) is going down. So, the appearance of Shapiro step within the resonance branch can be observed for certain range of spin-orbit coupling parameter, in our case at  $0 \le r \le 0.6$. For $r >0.6$, Josephson frequency $\omega_{J}$ is getting smaller than the radiation frequency  $\omega=0.485$ (due to the nonlinearity) and the maximum of Josephson frequency is getting outside of the locking region. I.e., the Josephson oscillations go out from the locking conditions, that's why there are no Shapiro steps at that $r$. We call this as a ``geometrical effect''.

\paragraph*{  Magnetization locking.}
The manifestation of the magnetic precession locking in the $m_{y}^{max}(I)$ dependence and its variation with $r$ is demonstrated in Fig.\ \ref 3(a), where the curves with the specific features are shown.
\begin{figure}[h!]
\centering
\includegraphics[width=4cm]{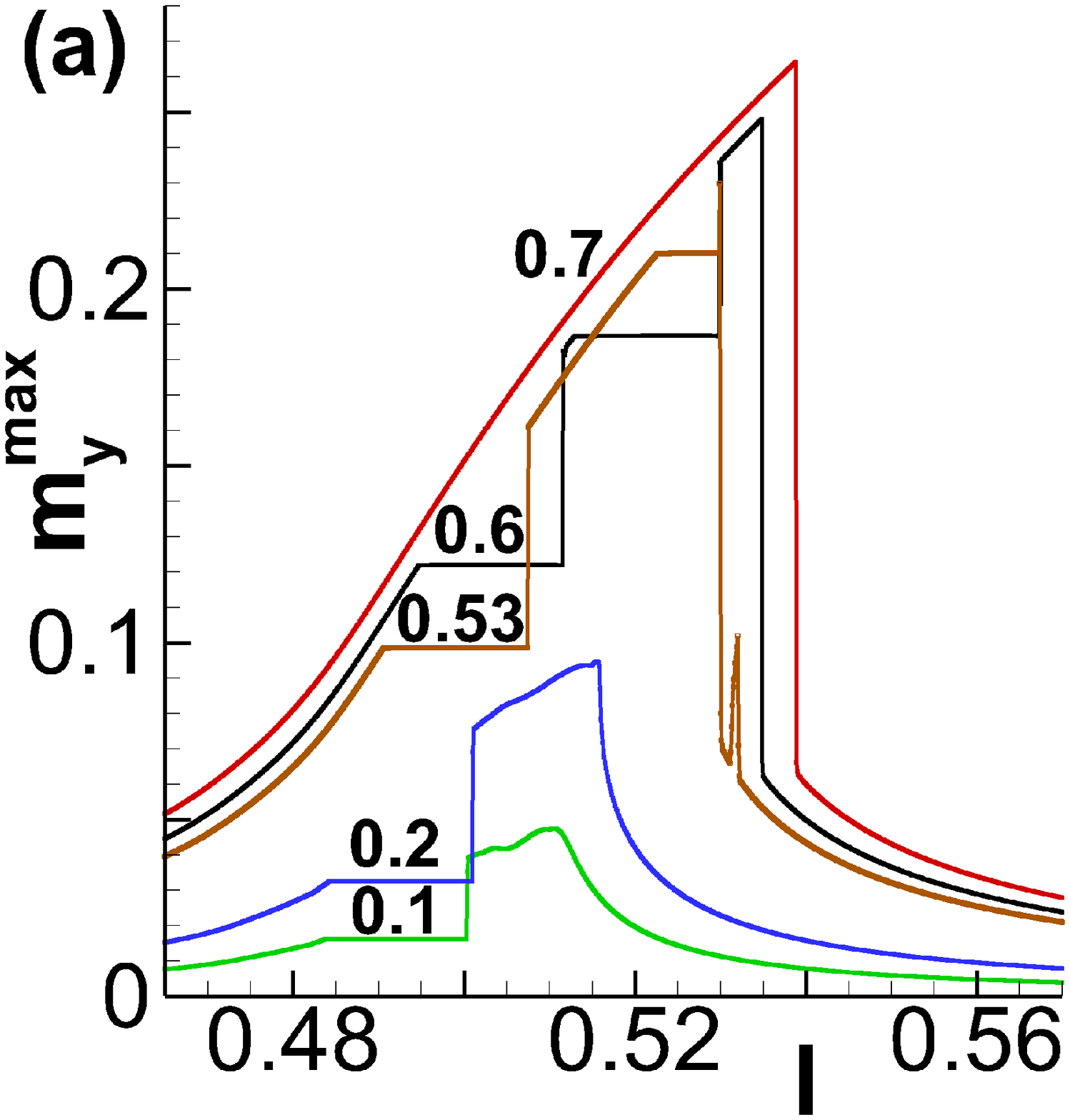}\includegraphics[width=4cm]{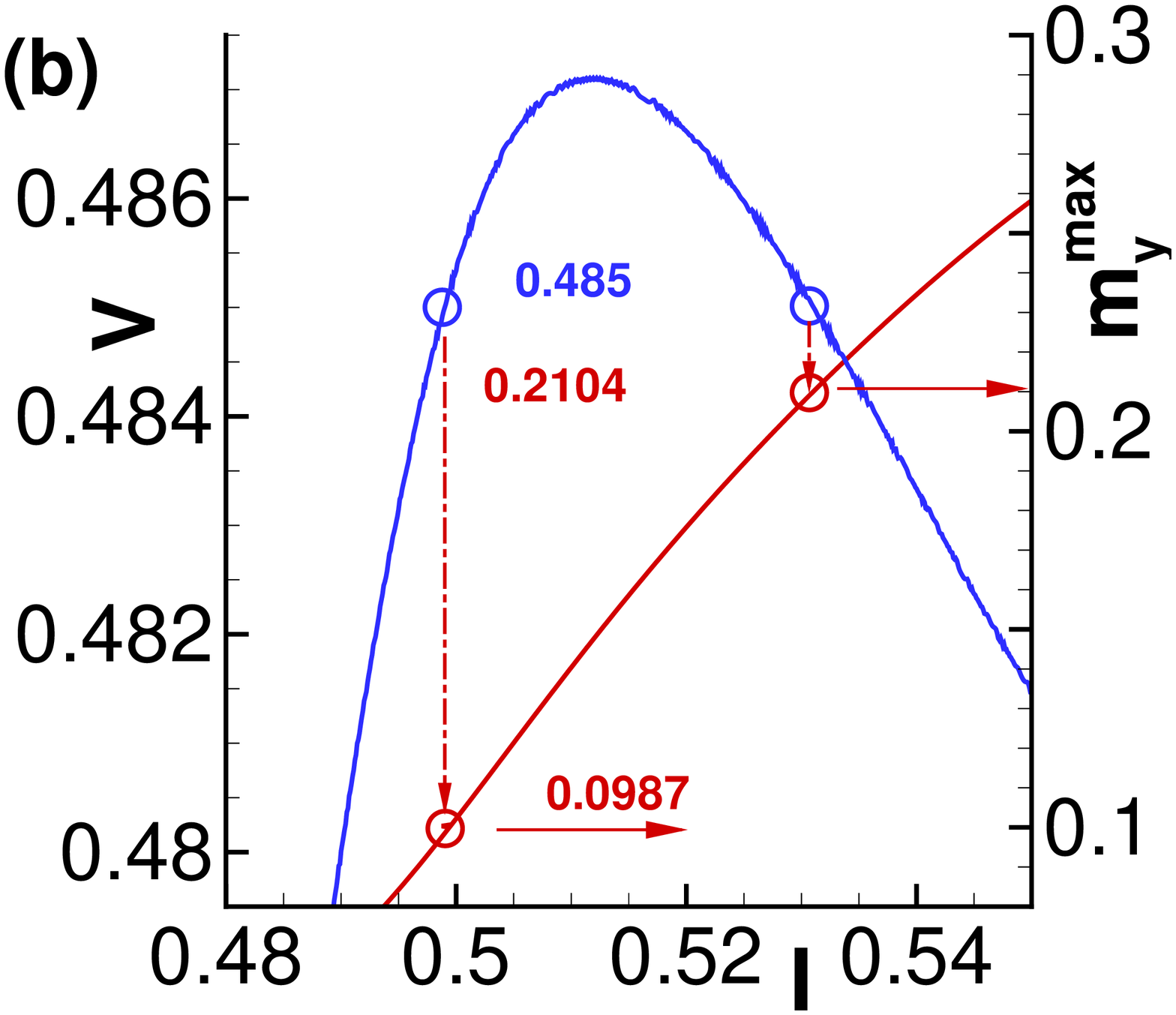}
\includegraphics[width=4cm]{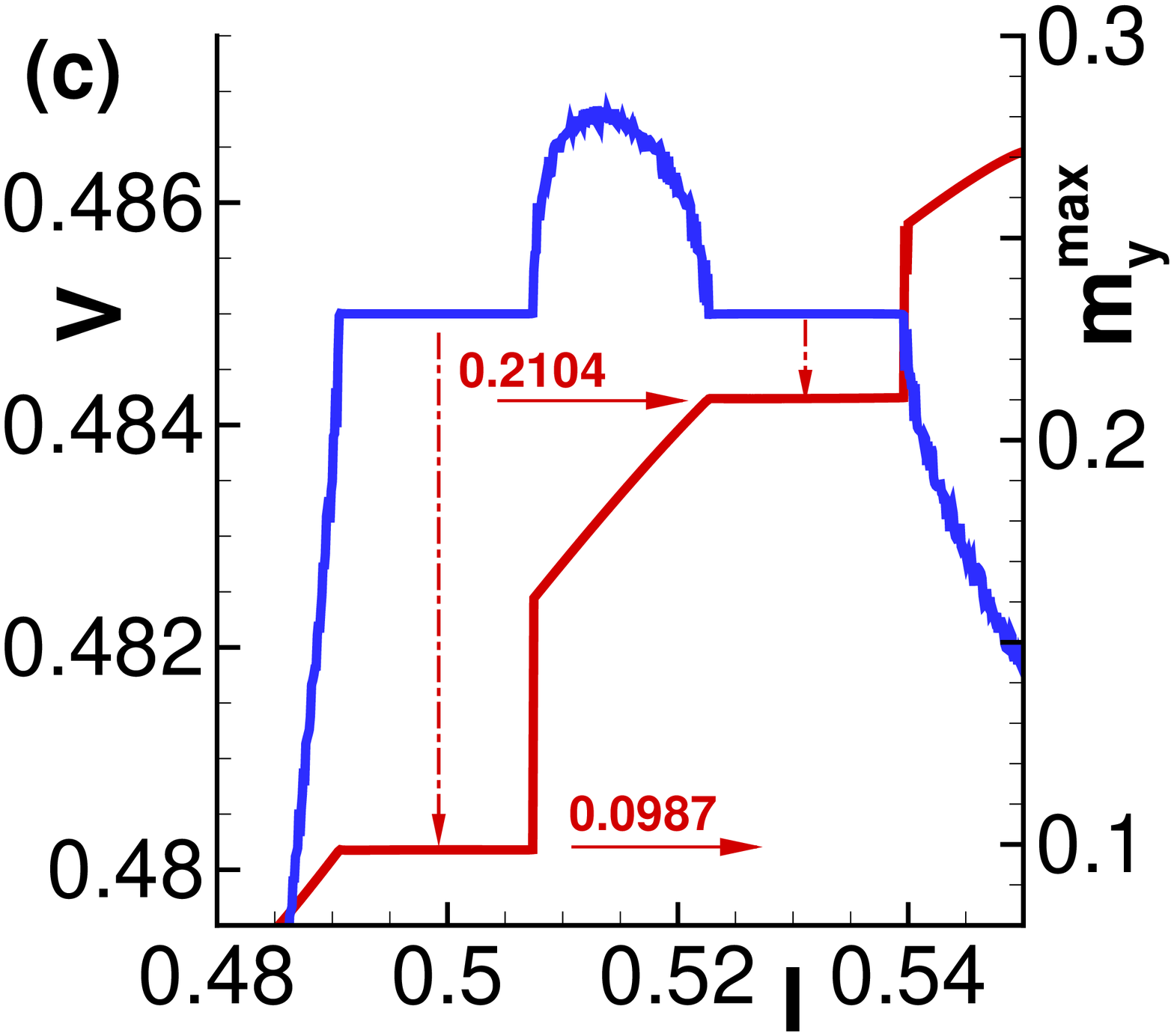}\includegraphics[width=4cm]{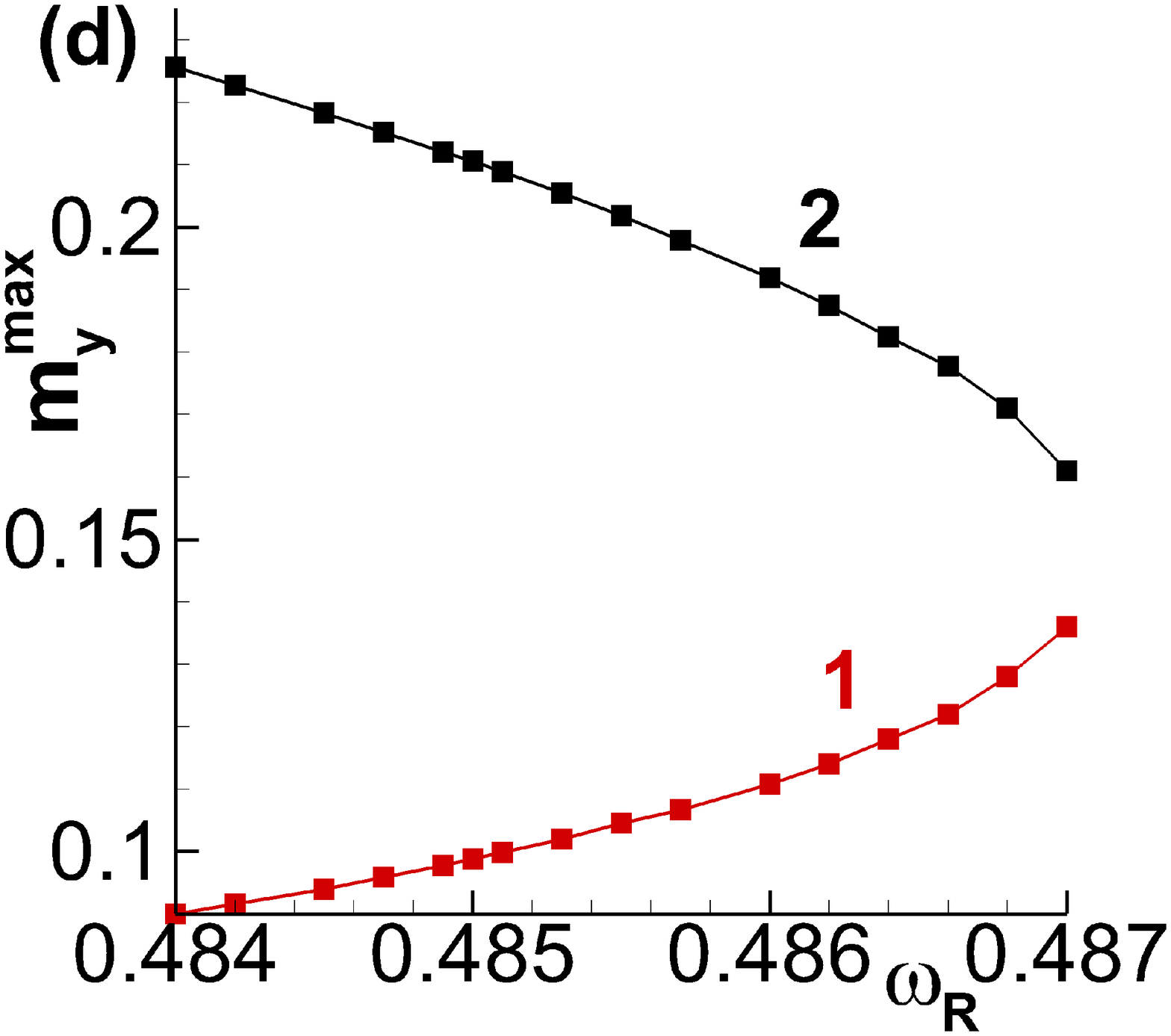}
\caption{(a) The bias current dependence of $m_{y}^{max}$ in the resonance region at different values of $r$ under radiation with $\omega_R=0.485$ and $A=0.1$;  (b) Procedure for locking steps determination (see text); (c) Results of numerical calculations of IV-characteristic and $m_{y}^{max}(I)$ dependence under radiation with the same parameters; (d) Position of the locking steps in $m^{max}_{y}$  as a function of $\omega_{R}$ at $r=0.53$ and $A=0.12$. Lines are  results of fitting by quadratic functions.  } \label{3}
\end{figure}
With an increase in the spin-orbit coupling in the interval $0.4<r<0.7$, the state with a NDR starts play an essential role. It is reflected by an appearance of the second step in $m_{y}^{max}(I)$ dependence that corresponds to the locking of magnetization precession at higher value of $m_{y}^{max}$. So, two locking steps with the different maximal magnetization  amplitude appear. This situation is demonstrated in figure by curves with $r = 0.53$ and $r = 6$.

A question appears about the position of the steps in $m_{y}^{max}(I)$ dependence, i.e., the value of  $m_{y}^{max}$ for the first and second steps. In difference with SS steps, when their  position in the IV-characteristics depends just on the frequency of the external radiation, the case of magnetization locking is more complex. Of course, $m_{y}^{max}$ also should be determined by the external radiation frequency. But in this case, it depends on the form resonance curve which depends on the parameters $G$, $r$ and $\alpha$. 

The procedure to find the position of the locking steps is demonstrated in  Fig.\ \ref 3(b). It shows the parts of the IV-curve and $m_{y}^{max}(I)$ dependence in the absence of the external radiation. The value of radiation frequency $\omega_R=0.485$ is marked by blue circles in the IV-curve in the states with positive and negative differential resistance. The bias current values in the IV-characteristics corresponded to the frequency of the external radiation determine the position of the locking steps in $m_{y}^{max}(I)$ dependence. The vertical lines which pass through these points and cross $m_{y}^{max}(I)$-curve,  fix the $m_{y}^{max}$ step positions at crossing points. In the discussed case, the first step is determined by value  $m_{y}^{max}=0.0987$, and the second one (in the state with NDR) - by $m_{y}^{max}=0.0.2104$. Figure \ \ref 3(c) shows results of direct numerical calculations of IV-characteristic and $m_{y}^{max}(I)$ dependence under radiation with $\omega_R=0.485$ and $A=0.1$ which demonstrates an agreement with proposed procedure.

The width of the steps is coincide with the width of SS steps in both cases.
From  Fig.\ \ref 3(b) it is clear that with an increase  in $\omega_R$ first step would go up, but the second one go down, because for second step, when blue circle go up, the red one go down. Results of detailed step's position calculation as a function of radiation frequency and results of their fittings by quadratic function  $a \omega_R^{2} + b \omega_R +  c$ with a=3623.1, b=-3504.2, c=847.37 for the first step and  a=-3455.9, b=3335.6, c=-804.65 for the second one, are presented in the Fig.\ \ref 3(d).

\begin{figure}[h]
		\centering		
\includegraphics[width=0.6\linewidth]{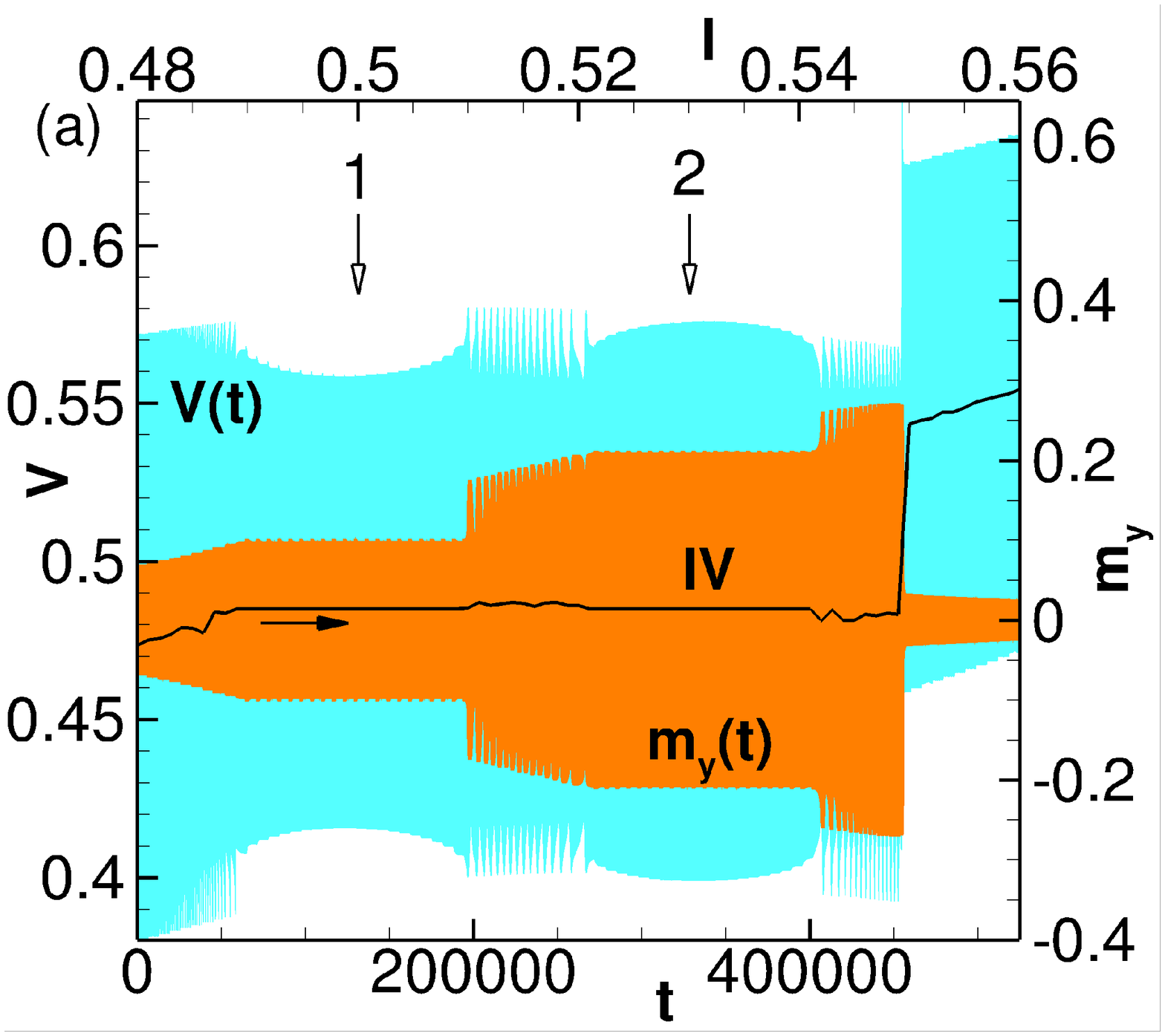}
\includegraphics[width=0.4\linewidth]{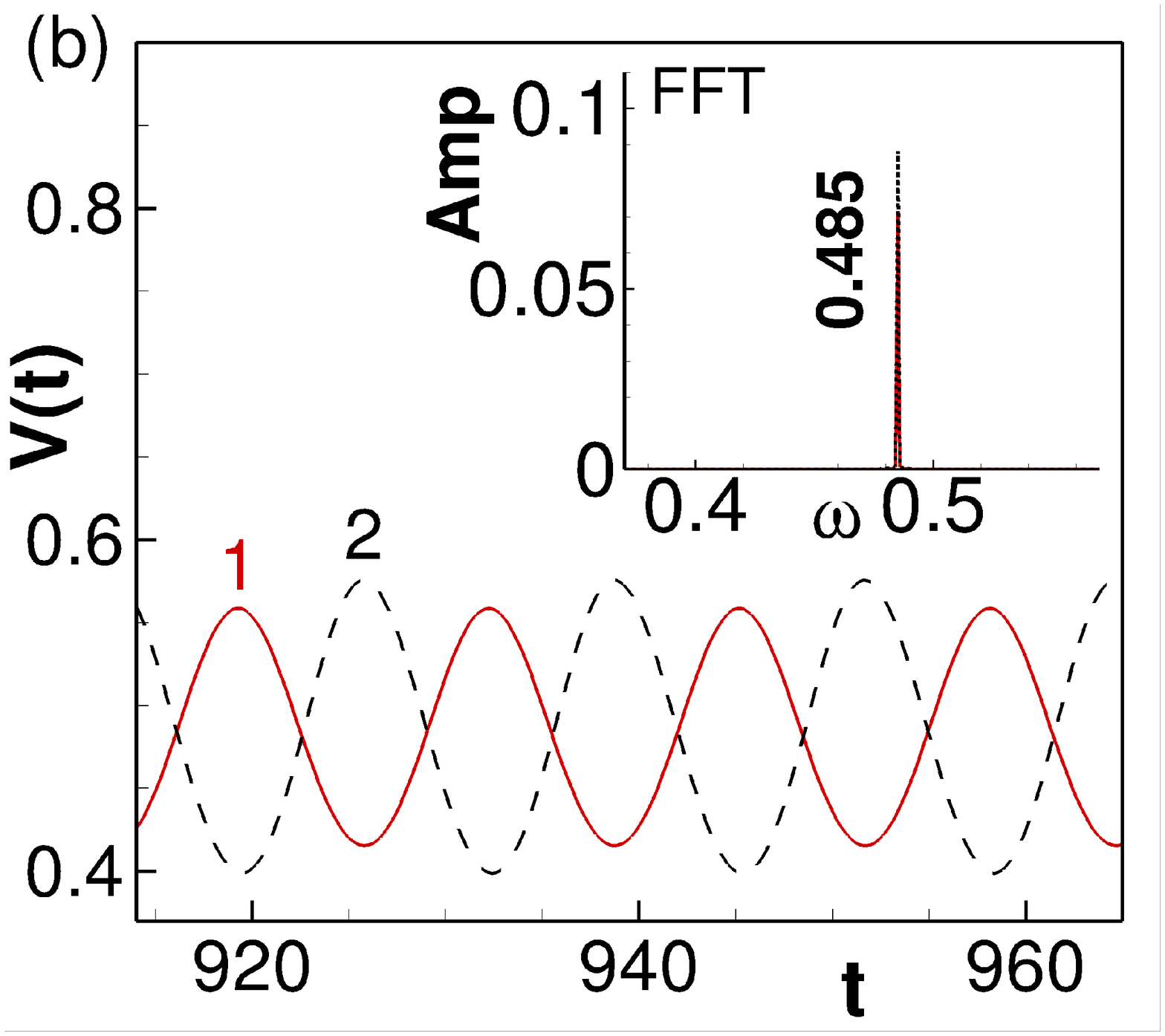}\includegraphics[width=0.4\linewidth]{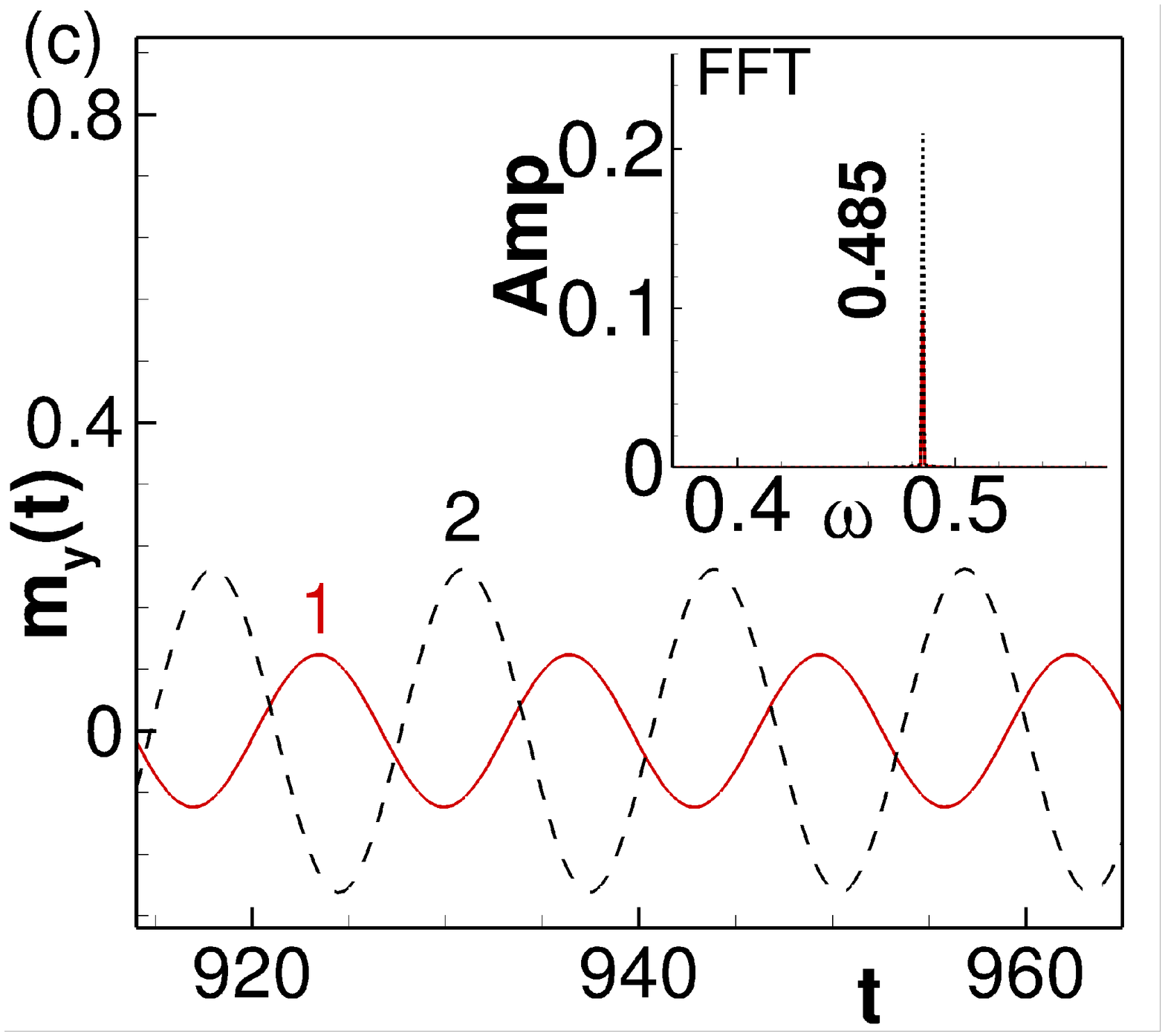}	\caption{(a) Enlarged resonance  area of IV-characteristic together with the time dependence of voltage $V(t)$ and magnetization $m_y(t)$ at $G=0.01$, $r=0.53$, $\alpha=0.01$, $A=0.12$ and $\Omega=0.485$ . The filled arrow denotes the direction of current change, the hollow arrows indicate the value of current at which the FFT has been performed. (b) Dynamics of voltage at current $I=0.5$ (indicated as $1$) and $I=0.53$ (indicated as $2$). The inset demonstrates the results of FFT analysis; (c) The same as (b) for $m_y$.}
		\label{4}
\end{figure}

\paragraph*{ Temporal dependence of $V(t)$ and $m_y$.}
An interesting effect can be seen in the time dependence of the voltage in the resonance  branch region, which is shown in Fig. ~\ref{4}(a). 

The amplitude $V(t)$ of the first SS decreases from the sides of the step to the its   middle. On the other hand, the amplitude $V(t)$ of the second SS increases from the sides of the step to the middle. This effect in the current-biased case is caused by the difference in the nonzero average supercurrent on the sides of the step, as well as the difference in the average voltage on the SS and the voltage without a step, which coincide in the step's middle  only. In the resonance  branch, in the positive differential resistance state, the supercurrent on the first step has a smaller value on the left side and larger on the right one. In the contrast, in the NDR state the supercurrent on the second step has the same behavior. This phenomenon appears because of the overlap of the FMR and locking conditions. Presented in Fig.~\ref{4}(a) time dependence of $m_y$ in the resonance  branch region  confirms the locking of the magnetization precession. The amplitude of oscillations of $m_y$ rise on the resonance  branch but in the SS region, the frequency and amplitude are fixed. So, the magnetization at the second step oscillates with the $\omega_R$ but the amplitude of oscillations larger than at the first step.

The enlarged time dependence of voltage in both SS regions is demonstrated in Fig.~\ref{4}(b). Due to the nonsynchronized nonlinear region between the steps, the oscillations are shifted in phase. As we mentioned already, the amplitude of oscillations are different because the second  step is in the part of the resonance branch with  NDR. The fast Fourier transform (FFT) analysis presented in the inset to Fig.~\ref{4}(b) demonstrates the locking of the voltage oscillations. The enlarged time dependence of $m_y$ component in the both locking states is demonstrated in Fig.~\ref{4}(c). The magnetic moment precesses with the same frequency but with different amplitudes. The FFT analysis of $m_y$ time dependence in the locking region is presented in the inset to Fig.~\ref{4}(c). It clearly shows one frequency $\omega_R$ only  which confirms the locking of the magnetization precession by the external periodic drive.

\paragraph*{Summary}
A method for controlling the dynamics of magnetization in the Josephson $\varphi_0$ junction is proposed. The possibility of locking the precession of magnetization by Josephson oscillations under  external electromagnetic radiation is demonstrated. An additional locking step  appears in a state with a negative differential resistance. The locking steps are determined by the frequency of  external radiation and by the form of the FMR curve, which depends on the ratio of Josephson to magnetic energy, spin-orbit coupling and Gilbert damping. Width of the locking steps is determined by amplitude of radiation and coincides with the width of Shapiro steps in the IV-characteristic. We have shown that external electromagnetic radiation can control not only the frequency of the magnetic precession, but also its amplitude.

The experimental testing of our results would involve SFS structures with ferromagnetic material having enough strong spin-orbit coupling. Using superconductor-ferromagnetic insulator-superconductor on a 3D topological insulator might be a way to have the spin-orbit coupling needed for $\varphi_0$ JJ \cite{bobkova-prb20}. The interaction between the Josephson current and magnetization is determined by the ratio of the Josephson to the magnetic anisotropy energy $\displaystyle G= E_{J}/(K \mathcal{V})$ and spin-orbit interaction $r$. The value of the Rashba-type parameter $r$ in a permalloy doped with $Pt$ \cite{hrabec-prb16} and in the ferromagnets without inversion symmetry, like $MnSi$ or $FeGe$, is usually estimated to be in the range $0.1-1$. The value of the product $Gr$ in the material with weak magnetic anisotropy $K \sim 4\times 10^{-5} KA^{-3}$ \cite{rusanov-prl04}, and a junction with a relatively high critical current density of $(3 \times 10^5 - 5 \times 10^6) A/cm^2$ \cite{robinson-sr12} is in the range $1-100$. It gives the set of ferromagnetic layer parameters  that make it possible to reach the values used in our numerical calculations for the possible experimental observation of the predicted effect.

\paragraph*{Acknowledgements.} The authors are grateful to I.R. Rahmonov for fruitful discussion of the results of this paper. The study was carried out within the framework
of the Egypt-JINR research projects.  Numerical simulations were funded by Project No. 18-71-10095 of the Russian Science Foundation.

\end{document}